\documentclass[structabstract]{aa}  
\usepackage{graphicx}
\usepackage{txfonts}
\usepackage{textcomp}

\usepackage{natbib}
\bibpunct{(}{)}{;}{a}{}{,} 

\catcode`\@=11
\def\gsim{\ifmmode{\mathrel{\mathpalette\@versim>}}
    \else{$\mathrel{\mathpalette\@versim>}$}\fi}
\def\lsim{\ifmmode{\mathrel{\mathpalette\@versim<}}
    \else{$\mathrel{\mathpalette\@versim<}$}\fi}
\def\@versim#1#2{\lower 2.9truept \vbox{\baselineskip 0pt \lineskip
    0.5truept \ialign{$\m@th#1\hfil##\hfil$\crcr#2\crcr\sim\crcr}}}
\catcode`\@=12

\begin{document}
   \title{The Calcium Triplet metallicity calibration for galactic bulge stars.
   \thanks{Based on observations taken with ESO telescopes at 
   the La Silla Paranal Observatory under programme ID 385.B-0735(B).}}

\author{
S. V\'asquez\inst{1,2} \and 
M. Zoccali\inst{1,2}   \and
V. Hill\inst{3}        \and
O. A. Gonzalez\inst{4} \and
I. Saviane\inst{4}     \and
M. Rejkuba\inst{5}     \and
G. Battaglia\inst{6,7}     	       
}

\institute{
Instituto de Astrof\'isica, Pontificia Universidad  Cat\'olica de Chile, 
Av. Vicu\~na Mackenna 4860, 782-0436 Macul, Santiago, Chile\\
\email{svasquez@astro.puc.cl}
\and
Millennium Institute of Astrophysics, Av. Vicu\~na Mackenna 4860, 
782-0436 Macul, Santiago, Chile
\and
Laboratoire Lagrange (UMR7293), Universit\'{e} de Nice Sophia Antipolis, 
CNRS, Observatoire de la C\^{o}te 
d'Azur, CS34229, 06304, Nice Cedex 04, France
\and
European Southern Observatory, Av. Alonso de Cordova 3107, Casilla 19, 
19001, Santiago, Chile
\and
European Southern Observatory, Karl-Schwarzschild Strasse 2, D-85748 
Garching, Germany
\and
Instituto de Astrof\'isica de Canarias, calle via L\'actea s/n, 38205 San 
Cristobal de La Laguna, Tenerife, Spain
\and 
Universidad de La Laguna, Dpto. Astrof\'isica, 38206 La Laguna, Tenerife, Spain
}


 
  \abstract
   {}
{We present  a new  calibration of the  Calcium II  Triplet equivalent
  widths  versus  [Fe/H], constructed upon K giant stars  in  the
  Galactic  bulge.   This calibration  will  be  used to  derive  iron
  abundances for the targets of the GIBS survey, and in general it is
  especially suited for solar and  supersolar metallicity giants, typical 
  of external massive galaxies.}
{About 150 bulge K giants  were observed with the GIRAFFE spectrograph
  at VLT, both at resolution R$\sim$20,000 and at R$\sim$6,000. In the
  first case, the spectra allowed us  to perform direct determination of 
  Fe abundances from several unblended Fe lines, deriving what we call 
  here high resolution [Fe/H] measurements. The low resolution spectra  
  allowed us to measure  equivalent widths of
  the two strongest lines of the near infrared Calcium II triplet at
  8542 and 8662 \AA.}
{By  comparing the  two  measurements we  derived  a relation  between
  Calcium  equivalent  widths  and  [Fe/H] that  is  linear  over  the
  metallicity range probed here, $-1<$[Fe/H]$<+0.7$. By adding a small
  second order correction, based on literature globular cluster data, we 
  derived the unique calibration equation
  $\mathrm{[Fe/H]_{CaT}} = -3.150 + 0.432W^{\prime} + 0.006W^{\prime 2}$, 
  with a rms dispersion of $0.197$ dex, valid across the whole metallicity
  range $-2.3<$[Fe/H]$<+0.7$.}
{}

   \keywords{ Stars: abundances -- Galaxy: bulge -- Techniques: spectroscopic
               }

   \maketitle
%

\section{Introduction}
	
The Calcium II  Triplet (CaT) at $\sim$8500 \AA~ is one  of the most widely
used metallicity  index, as  well as an  excellent feature  to measure
radial  velocity  at low  spectral  resolution.   The three  lines  at
$\lambda$8498,  $\lambda$8542, $\lambda$8662  \AA~ are  so strong  that
they can  be measured easily at  low resolution and at  relatively low
signal to noise. In addition, their location in the near-infrared part of
the spectrum  is ideal to observe  the brightest stars of  any not too
young  stellar population,  namely cool  giants. CaT  spectra of  cool
giants can be obtained with reasonable exposure time both for external
galaxies in the local group, too far away to be observed at high 
spectral resolution, and
for Milky Way  stars in high extinction regions, such  as the Galactic
bulge.

Obviously the  popularity of the  CaT spectral feature resides  in how
accurately it can be used to measure metallicities. \cite{armandroff+1988}
first demonstrated  that the  equivalent widths  (EWs) of  CaT
lines, in the integrated spectra  of globular clusters (GCs), strongly
correlated with  the cluster metallicity  [Fe/H].  A few  years later,
\cite{olszewski+1991}   and    \cite{armandroff+1991}   analyzed   the
behaviour of CaT lines in  individual cluster stars. They noticed that
the EWs  of CaT lines show  a dependence not only  on metallicity, but
also  on absolute  magnitude.  They  therefore introduced  the use  of
``reduced equivalenth widths''  ($W^\prime$), which corresponds to the sum 
of some combination of the individual EWs, weighed by
the difference between the star $V$ magnitude and the magnitude of the
Horizontal Branch in the same cluster ($V-V_{\rm HB}$).

Several empirical  relations between the reduced EWs
of  CaT lines  and  the  [Fe/H] abundance are present in the literature. 
Most  of  them used  star
clusters, for which [Fe/H] abundance could be derived in several ways,
and not  necessarily for the  same stars  for which CaT  was measured.
Among   those,   a   very   comprehensive   one   is   the   work   of
\cite{rutledge+1997} who derived CaT metallicities for 52 Galactic GCs
in  a homogeneous  scale,  covering the  range $-2<$[Fe/H]$<-0.7$.  By
comparing  their  scale  with the  classical  \cite{ZW84}  metallicity
scale, they found a non-linear  relation between the two.  Conversely,
a  linear  relation  was  found   between  the  CaT  metallicities  by
\cite{rutledge+1997}  and the  metallicities  derived by  \cite{CG97}.
The latter are based on Fe lines, measured on high resolution spectra.

Traditionally, the  CaT metallicity  calibration has  been constructed
based on  RGB stars  in globular  \citep[e.g.]{cole+2004, warren+2009,
  saviane+2012}  or  open  clusters  \citep{carrera+2007,carrera+2013}
Open clusters allowed  Carrera et al. to extend  the metallicity range
up to [Fe/H]$\sim$+0.5, at the same  time probing a younger age regime
(13  Gyr$<$age$<$0.25  Gyr).  The  CaT  EWs  seem  to be  only  weakly
dependent on  the age of the  star, although only a  few star clusters
constrain  the  relation at  high  metallicity,  and anyway  old  star
clusters at supersolar metallicities are not available to set a robust
constrain on the age dependance.

\cite{battaglia+2008} compared a  CaT calibration based on GCs with
one based on red giants in  dwarf spheroidal galaxies.  The latter are
complex  stellar   populations  with  a   range  of  ages   and,  most
importantly, a  range of [Ca/Fe]  abundance ratio. They  conclude that
the two  relations are fairly  consistent within the  errors, although
small differences  exist. In  particular, a  CaT-to-[Fe/H] calibration
based on  GCs would overestimate  the [Fe/H] of low  metallicity stars
([Fe/H]$<-2.2$) by $\sim$0.1 dex,  while underestimating the [Fe/H] of
(relatively) high metallicity  stars ([Fe/H]$>-1.2$) by $\sim$0.1--0.2
dex.  It is  important to emphasize that their analysis  is limited to
the metallicity range $-2.5<$[Fe/H]$<-0.5$.

More recently \cite{starkenburg+2010} presented  a synthetic spectral
analysis of the  CaT method to derive iron  abundance, with particular
emphasis at the low metallicity regime. They derive a new calibration,
valid in the range $-4<$[Fe/H]$<-0.5$, that deviates from the commonly
used linear relation for [Fe/H]$<-2$.

To  summarize, CaT  to [Fe/H]  calibrations at  solar and  super-solar
metallicities are very sparse, and  based only of young open clusters.
Deviations from linearity, in the  relation between CaT EWs and [Fe/H]
in the high metallicity regime would not be surprising, given that the
Calcium  lines   are  highly   saturated  already  well   below  solar
metallicity.   Departures from  local thermodynamical  equilibrium are
also known  to affect  the core  of the  lines \citep{mashonkina+2007}
because of  saturation.  In this  paper we probe the  high metallicity
regime of this relation based on old red giants in the Galactic bulge.

This  study  is the  result  of  a pilot  project  we  carried out  to
investigate the  possibility of deriving the  metallicity distribution
function (MDF) of the Galactic bulge,  based on  an iron abundance  
scale, by  measuring CaT
lines in low/intermediate resolution GIRAFFE  spectra. To this end, we
acquired both high (R$\sim$20,000)  and low (R$\sim$6,000) resolutions
GIRAFFE spectra for  a large sample of  K giants in two  fields in the
Galactic  bulge. When  preliminary results  were very  encouraging, we
were  granted an  ESO  Large Programme  (187.B-0909;  PI: Zoccali)  to
observe 200-400  red clump (RC)  giants in each of  25 fields in  the Galactic
bulge, with GIRAFFE  at low resolution. Those  were then combined
with other archive  data obtained with the same  instrument and setup,
also  targetting bulge  RC  stars.  The  resulting dataset,  including
spectra for  6392 stars in  31 fields, is  what we called  the Giraffe
Inner   Bulge   Survey   (GIBS),   described   in   \cite[][hereafter
  Paper~I]{zoccali+2014}. Paper~I  also lists  the radial  velocity of
individual  stars in  the  sample, and  presents  radial velocity  and
velocity dispersion maps of the inner bulge.

We focus here on the CaT to [Fe/H] calibration, that will be used in a 
forthcoming paper to derive the MDF in 25 fields in the Galactic bulge.


\section{Observation and data reduction}
	
\begin{figure}
\centering
    {\includegraphics[angle=0,width=1.0\columnwidth]{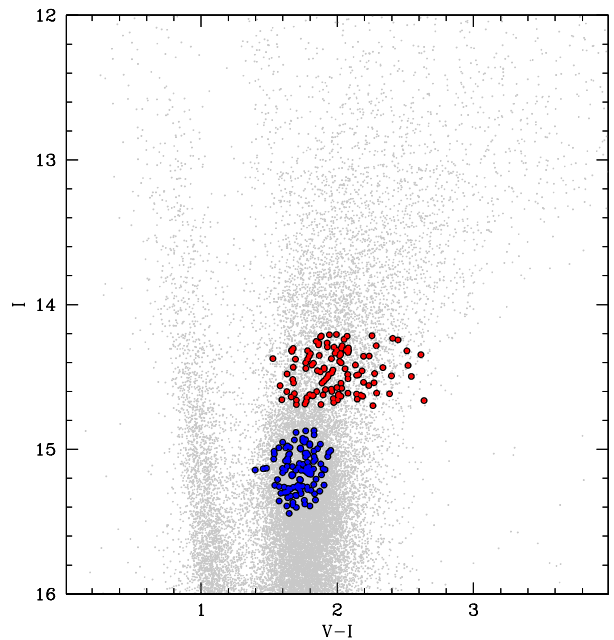}}
    \caption{The target selection is shown over the V and I band colour 
    magnitude diagram using color dots. Red and blues correspond to the
    RC and RGB stars from Baade's window ($b=-4$).}
    \label{fig:1_cmd}
\end{figure}
	
The target  stars we  used to  construct the  CaT calibration  are all
located in Baade's  Window, and they consist of two  samples: 80 stars
in the bulge RC, and 68 stars  on the RGB, $\sim$0.7 mag brigther than
the RC. The medium resolution  spectra analysed here
were taken with the FLAMES-GIRAFFE multifibre spectrograph at the VLT,
in  Medusa  mode,  within  the  ESO  program  ID  385.B-0735(B).   The
spectrograph  allows  us  to  obtain spectra  for  $\sim$130  objects,
including sky positions, over a 25  arcmin diameter field of view. We
selected the LR08 setup, which covers the wavelength range
between  8206  and  9400  \AA,  with a  spectral  resolution  of  $\rm
R\approx6500$. The exposure time needed to reach a mean signal to noise 
of $\sim 50$/pix were 900  and 1800 seconds for RGB and RC 
stars, respectively.

High  resolution  spectra  for  the targets  in  Baade's  Window  were
obtained  from Zoccali et al. (2008) and Hill et al. (2011) who used 
the  same instrument  in the  setups HR13  and HR14,  at
resolution    $\rm   R\approx22,500$    and   $\rm    R\approx28,800$,
respectively. Figure \ref{fig:1_cmd}  shows the corresponding color 
magnitude diagrams (CMDs) for  the target selection.

\begin{figure}
\centering
    {\includegraphics[angle=0,width=1.0\columnwidth]{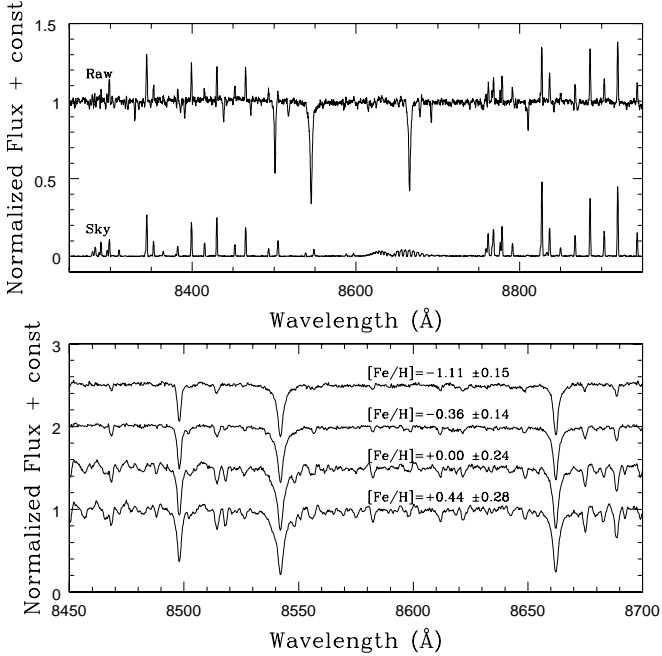}}
    \caption{The upper panel shows the normalized raw spectrum for a typical 
    Milky Way bulge star, together with the corresponding sky spectrum. 
    The lower panel shows processed spectra for 4 bulge stars spanning the 
    typical metallicity range of the Galactic bulge. The increasing strength 
    of the CaT lines with metallicity is evident, together with the increasing
    number of small atomic lines contaminating the continuum.}
    \label{fig:2_spec}
\end{figure}

The data reduction was done  using the instrument pipeline provided by
ESO  (version  2.8),  which  performs bias,  flat  field  corrections,
wavelength  calibration,  and  1-D  extraction based  on  the  daytime
calibration images. Since the pipeline does not include the sky 
subtraction routine, we performed   this task with
\textsc{IRAF} as follows.  As a first  step, a master sky spectrum was
created by  combining all the  sky spectra ($\sim20$ per  setup) using
the median  and a  sigma clipping algorithm.   Figure \ref{fig:2_spec}
(upper panel) shows the raw spectrum of a typical star, around the CaT
region,  together   with  the  master   sky  spectrum  for   the  same
exposure. It  is clear  that the  three CaT lines  do not  suffer from
strong contamination from sky emission  lines compared with CaT line 
strength. The master sky spectrum
was then  subtracted from each  individual science spectrum  using the
\textsc{IRAF} task  \texttt{skytweak}. The latter allows  shifting and
scaling  of  the master  sky  spectrum  until  a good  subtraction  is
achieved. In order to find the best shift and scale factor we used the
strong emission lines located in  the wavelength range between 8261 to
8481 \AA~ and 8739 to 9071 \AA.

The   spectra   were   normalized    with   the   \textsc{IRAF}   task
\texttt{continuum}, using the wavelength  range between 8250-8461 \AA,
8568-8650 \AA,  and 8700-9273  \AA, in  order to  avoid the  CaT lines.
Finally, radial velocities were measured by means of cross-correlation
(\texttt{fxcor}  task) against  a  template synthetic  spectrum for  a
typical metal poor K giant star ($T_{\rm eff}=4750$, $\log\;g=2.5$ and
$\rm [Fe/H]=-1.3$). Such low metallicity was chosen in order to ensure
that the CaT lines in the  template were relatively deep but free from
contaminations by other lines on their wings.
	  
Example  of  the  final   spectra,  sky  subtracted,  radial  velocity
corrected and  continuum normalized  are shown in  the lower  panel of
Fig.~\ref{fig:2_spec},  for  stars   of  different  metallicities.  An
arbitrary shift has been applied along the y-axis, to avoid overlaps.

\section{The Calcium Triplet calibration}

The spectral  region around the  CaT feature (8350-9000  \AA) includes
several  lines which  might dominate  the spectrum,  depending on  the
spectral type  of the star under  analysis. \cite{cenarro+2001} showed
that stars with  spectral type between F5 ($T_{\rm eff}\sim6500$ K) and 
M2 ($T_{\rm eff}\sim3700$ K)  are the best targets
to measure the CaT lines, due  to their only marginal contamination by
FeI, MgI  and TiI lines.  CaT lines in  hotter stars are  blended with
Paschen lines, while  in cooler stars the continuum is  plagued by TiO
molecular bands. Being of K spectral type, our targets are well inside
the recommended temperature range.

Additional constraints have  been proposed in terms  of the brightness
range where  a single gravity correction  can be used (see  section 3.2
for   details).   \cite{carrera+2007}   defined  a   minimum  absolute
magnitude    $M_{V}\leq    1.25$    or    $M_{I}\leq    0.0$,    while
\cite{dacosta+2009} recommends a lower  brightness limit of $\sim 0.2$
mag  below the  RC (or  the horizontal  branch). Both  our RGB  and RC
targets are inside  these constraints, with the RC  stars being within
the intrinsic  dispersion --  mainly due  to metallicity  and distance
spread -- of the RC in Baade's Window.

\subsection{The equivalent width measurements}

\begin{figure}
\centering
    {\includegraphics[angle=0,width=1.0\columnwidth]{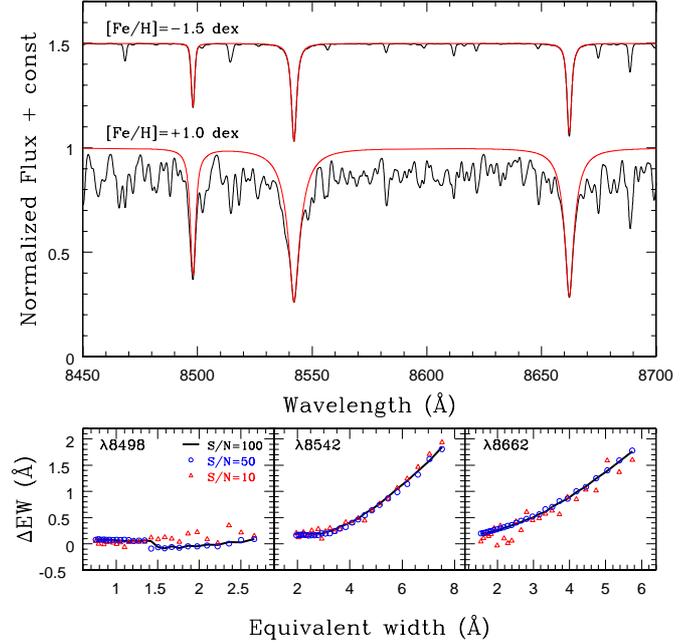}}
    \caption{The upper panel shows for comparison the more metal poor and 
    metal rich synthetic
    spectra from our library, in the region around the CaT lines. 
    In both cases, the black 
    line shows the spectra including the lines of all elements, while the
    red lines shows only the three CaT lines. The lower panels show the 
    difference between the {\it real} EW, as measured in the red spectra, 
    and the {\it measured} EW, as measured in the black spectra, for each
    of the three CaT lines, as a function of EW and for three values of S/N.}
    \label{fig:3_synth}
\end{figure}

While both  our RC  and RGB  target stars  are inside  the recommended
temperature and  gravity range  for CaT measurements,  the metallicity
distribution of  the Galactic bulge reaches (extends) to  significantly
higher metallicity than typical star  clusters for which  this calibration
has been  defined and used  so far.  The  impact of the  metal content
over  the CaT  EWs  measurements  is mostly  driven  by the  continuum
depletion  due to  the increasing  contamination of  the CaT  lines by
other  atomic lines.   In order  to explore  the impact  of the  metal
content on  the CaT EWs measurement,  we analysed a grid  of synthetic
spectra generated using  {\it turbospectrum}. The latter is  a code of
line synthesis described in \cite{alvarez+1998} and improved along the
years by  B.  Plez.  In this  case it  was used together  with the  grid of
OSMARCS  spherical  model   atmospheres\footnote{models  available  at
  http://marcs.astro.uu.se/}  \citep{gustafsson+2008}.  The  synthetic
spectra  have  the  same  resolution as  the  GIRAFFE  spectra, and a 
set of stellar parameters representative of a K giant
star ($T_{\rm eff}=4750$ and $\log\;g=2.5$).  Only the metallicity was
allowed to  vary across  the range from  [Fe/H]=$-1.5$ to  the extreme
case of  [Fe/H]=+1.0 dex\footnote{The metallicity distribution  of the
  Milky Way bulge spans the range between $-1.5$ up to $+0.7$ dex}, keeping
[Ca/H]=[Fe/H] in the whole grid.

Synthetic spectra for stars at the two extreme ends of the metallicity
grid  are plotted  in the  upper panel  of Fig.\ref{fig:3_synth}.  The
black  line is  the  synthetic  spectrum containing  list  of all  the
element in  the wavelength range  between 8450  and 8700 \AA.  The red
line,  in contrast,  shows only  the  three CaT  lines. This  exercise
clearly shows  the depletion  of the continuum  (pseudo-continuum), at
high   metallicity,   due  to   the   presence   of  many   unresolved
lines.  Because   we  deal  with  normalized   spectra  (without  flux
calibration) we need  to include the effect of  continuum depletion in
the empirical calibration itself.

The measurements  of the CaT EWs  can be done by  fitting an empirical
function to each line profile  and calculate the corresponding EW from
direct  integration of  the  function. Different  functions have  been
explored  over  the  years.  \cite{armandroff+1991}  used  a  Gaussian
profile,       when       fitting        stars       in       globular
clusters. \cite{rutledge+1997}, instead, used  a Moffat function. Both
functions  give  good  results  only   for  stars  of  relatively  low
metallicity. In fact,  \cite{cole+2004} showed that in  the metal rich
regime, the strong wings of the CaT lines can be reproduced by the sum
of  a Gaussian  component, fitting  the  line core,  and a  Lorentzian
function,   fitting   the   wings.    \cite{erdelyi-Mendes+1991}   and
\cite{battaglia+2008}  showed that  the  EWs of  CaT  lines is  driven
mostly by  the wings,  because of the  strong pressure  broadening. In
other  words,  the strength  of  the  line  is  mostly driven  by  the
electronic  pressure (i.e.,  metallicity),  rather than  by the  usual
combination  of  temperature  and elemental  abundance,  shaping  weak
lines.

\begin{table}
\caption{}
\centering
\begin{tabular}{cc}
\hline 
Feature (\AA) & Line bandpass (\AA) \\ 
\hline 
8498.02 & 8491.02 -- 8505.02 \\ 
8542.09 & 8530.09 -- 8554.09 \\ 
8662.14 & 8653.14 -- 8671.14 \\ 
\hline 
\end{tabular} 
\label{table_line_band}
\end{table}

In order to test the accuracy  of the Gaussian+Lorentzian fit across a
range  of  metallicity   and  signal  to  noise,   we  performed  some
measurements on the  grid of synthetic spectra described  above.  As a
first step, EWs were measured in the synthetic spectra containing only
Calcium lines by means of numerical integration.   These were assumed  
to be the  {\it real} EWs.   As a
second step, the  grid of synthetic spectra was convolved to the 
resolution and resampled in wavelength to match real data
 and degraded  to different S/N, from 10 to 100
in  steps  of  10.  A  Poisson  noise  model  was  used for  the  last
operation.   The  resulting  spectrum  was  then  re-normalized  to  a
pseudo-continuum  of  1,  and  the  combination of  a  Gaussian  and  a
Lorentzian  function was  fitted  to the  CaT  lines. The  wavelength
ranges  used  by  \cite{armandroff+1988}  for the  fit  were  slightly
modified,  as  listed  in  Table \ref{table_line_band},  in  order  to
improve   the    fit   of   the    wings.    The   lower    panel   of
Fig.~\ref{fig:3_synth} shows  the difference between the  measured and
the {\it real} EWs, for each of the CaT lines, as a function of EW
(metallicity) and  S/N.  It is  clear that the  EWs of the  two strong
lines  would be  underestimated  in  the data,  and  more  so at  high
metallicities.  This  systematic  reaches  up to  $\sim$1  \AA~  around
[Fe/H]=+0.5. However, the trend is very  smooth and it is not affected
by  S/N,  down   to  a  S/N$\sim$10,  much  lower   than  our  poorest
spectrum. As a  consequence, a correction for this  systematic will be
implicitly  included   in  the  [Fe/H]-vs-CaT   empirical  calibration
equation derived in Section 3.4.  For the weakest line, the systematic
difference between  {\it real} and  {\it measured} EW is  much smaller
than  for  the   strong  lines.  However,  a  step   is  found  around
EW$_{1}=1.5$ \AA, due to the appearance of some blends in the wings of
this  line. Additionally,  the weakest  line is  the most  affected by
unperfect sky subtraction, due to the  presence of two sky lines close
to its wings. For this reason, we discarded this line in the following
analysis.

\begin{figure}
\centering
    {\includegraphics[angle=0,width=1.0\columnwidth]{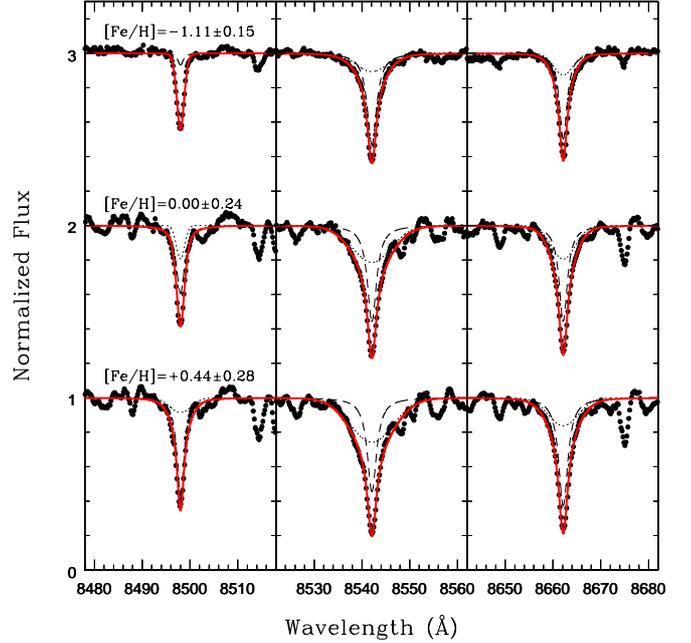}}
    \caption{Fitting examples for three bulge stars covering the metallicity 
    range of the Milky Way bulge. The red line correspond to the best fitting
    function found for each line, while as dashed lines are plotted the 
    individual contribution from Gaussian and Lorentzian functions.}
    \label{fig:4_ew_comp}
\end{figure}

\begin{figure}
\centering
    {\includegraphics[angle=0,width=1.0\columnwidth]{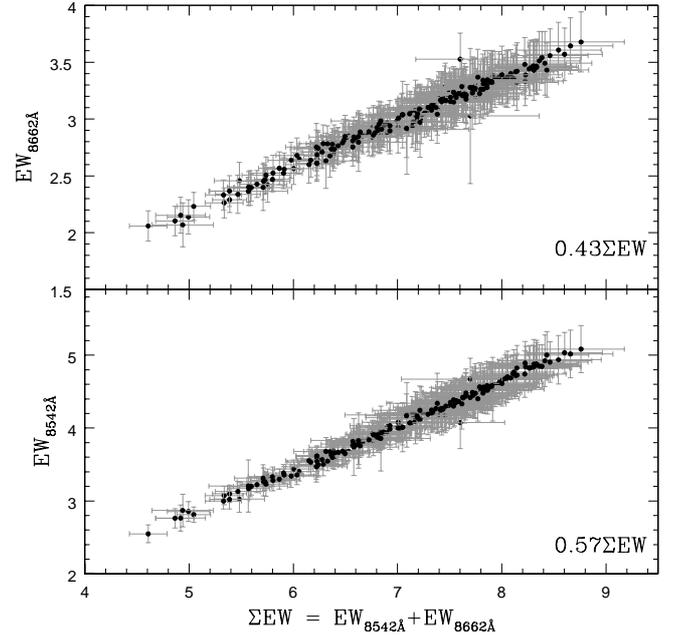}}
    \caption{The strength of the two strongest CaT lines versus the sum 
    of the two lines, $\Sigma~EW$. The corresponding fractional 
    contribution with respect to $\Sigma~EW$ is shown in each panel.}
    \label{fig:ew_ratio}
\end{figure}    	
    	
\begin{figure}
\centering
    {\includegraphics[angle=0,width=1.0\columnwidth]{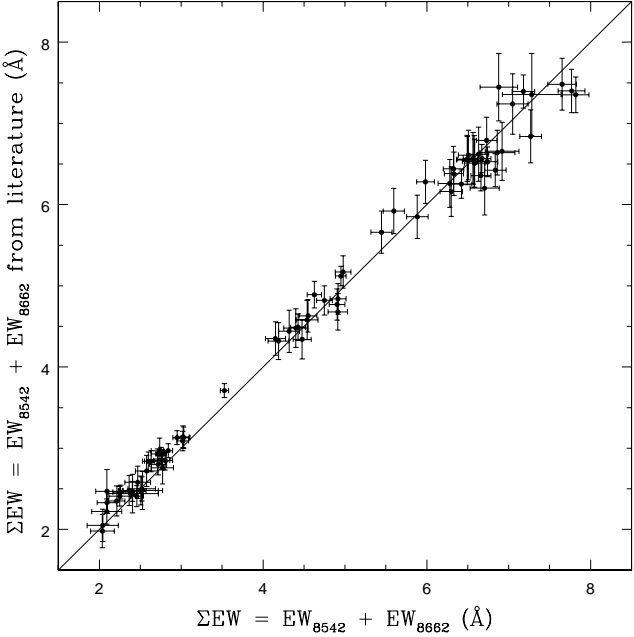}}
    \caption{Comparison between our measurements and the literature values of
    the sum of the two strongest lines for different stellar cluster taken 
    from \cite{warren+2009} and \cite{saviane+2012}.}
    \label{fig:5_comp_ew}
\end{figure}

Examples of line fitting are shown in Fig.~\ref{fig:4_ew_comp} for three
observed spectra covering the metallicity range of the Milky Way bulge.
For each star, the big black dots show the observed spectrum, the red 
line is the final fit, while the thin dotted and dashed lines show the 
individual contribution of the Gaussian and the Lorentzian profiles, 
respectively. Additionally, the line-ratio between the two strongest CaT
lines is remarkably constant across the metallicity range spanned by 
the bulge data. Figure~\ref{fig:ew_ratio} shows the 
contribution of the two strongest individual CaT lines to the sum of the
two. The data can be fitted by a single line, with the slope indicated in
the label. This line ratio can be used, for instance, to predict the EW 
of one of the lines, when the other is not available for a direct measurement.
It can also be used to test the consistency of the measurements, in order
to reject lines with anomalous line ratio. This is particularly useful when
dealing with large surveys.

As will be explained in the following sections, we will use cluster stars to 
constrain the final metallicity CaT calibration. Before doing it, we need
to verify that our EWs measurements are in the same scale as those used 
in previous works. To this end, we obtained the spectra of the two most
metal rich clusters (NGC~6791 and NGC~6819) in the sample by \cite{warren+2009} 
and re-measure EWs. The spectra kindly provided to us, by private communication, 
were already wave-calibrated and sky-subtracted, therefore we only applied the 
same continuum normalization procedure and EW measurement used for our data. 
Additionally, in order to to probe a wider range of metallicity, we also measured 
a sample of globular cluster stars, drawn from \cite{saviane+2012},
spanning the metallicity range $-2.3<$[Fe/H]$<+0.07$. The selected stars
belong to the globular clusters 
M 15, NGC 6397, M4, NGC 6553, and NGC 6528. Figure~\ref{fig:5_comp_ew}
shows the  excellent agreement between our measurements and the literature
one, for the  sum of the EWs of the two
strongest CaT lines, hereafter $\rm \Sigma EW$ ($=  EW_{8542} +  EW_{8662}$).

\subsection{The reduced equivalent width}
	
\begin{figure*}
\centering
{\includegraphics[angle=0,width=1.0\columnwidth]{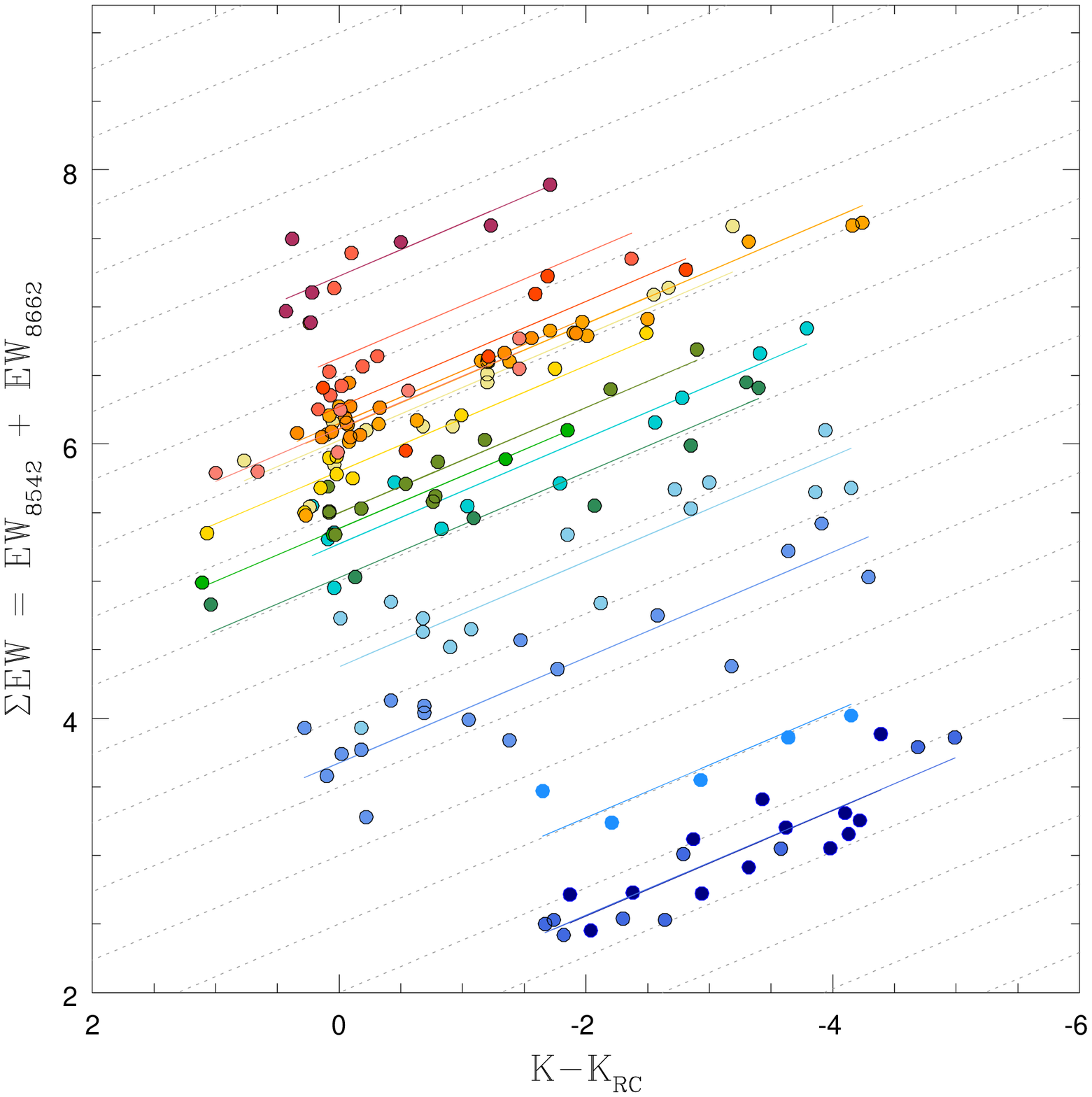}}       
{\includegraphics[angle=0,width=1.0\columnwidth]{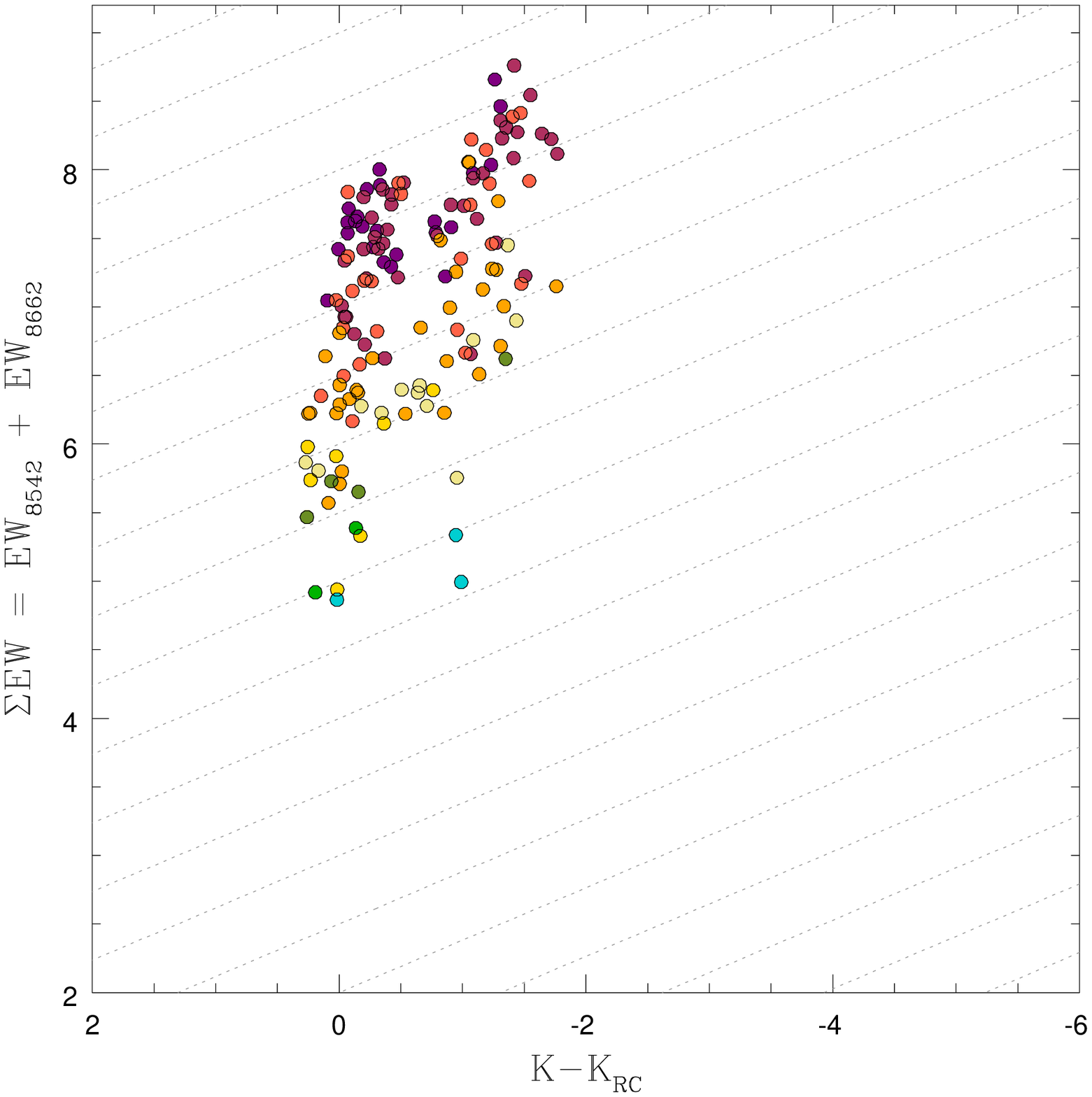}}         
\caption{The CaT equivalent widths of the star cluster RGBs (left panel) and 
Galactic bulge stars (right panel) are compared with their corresponding 
magnitude above the mean red clump magnitude. The colour code is used to 
mark clusters of different metallicity, together with their best-fitting 
lines, assuming the best common slope of $0.384$ \AA\~ mag$^{-1}$. Thin
dotted lines in each panel show this slope, to guide the eye.}
\label{fig:6_7_beta_clusters}
\end{figure*}

The strength of an atomic line, in general, is driven by the elemental
abundance, the effective  temperature, and the surface  gravity of the
star.   Both \cite{armandroff+1991}  and \cite{olszewski+1991}  showed
that the combined effect of  effective temperature and surface gravity
on  the  EWs  of CaT  lines  can  be  parametrized  as a  function  of
luminosity alone.   The CaT  EWs corrected by  this effect  (i.e., the
{\it reduced}  EWs) depend only upon  metallicity.  Specifically, they
found that, in  each given cluster, red giant stars  follow a straight
line in the luminosity-$\rm \Sigma EW$ plane. The slope of the line is
identical for different clusters, while  its zero point depends on the
cluster metallicity. As  a proxy for luminosity, some  authors use the
absolute magnitude of the star \citep{carrera+2007, carrera+2013}, or,
most commonly, the difference between  the star apparent magnitude and
the  magnitude  of  the  cluster   horizontal  branch,  or  red  clump
\citep{armandroff+1991,   rutledge+1997,  cole+2004,   battaglia+2008,
  warren+2009, saviane+2012, mauro+2014}. We adopted the second option,
in order  to minimize the  uncertainty in the distance  and reddening,
both variable across the Galactic bulge.
 
The $K_{s}$  magnitude, from the VVV  survey \citep{minniti+2010}, was
used  for our  bulge stars.  In order  to determine  the slope  in the
luminosity-$\rm \Sigma EW$  plane we used the star  cluster data from
\cite{warren+2009}. For  each
cluster  star we  plot $\rm  \Sigma EW$  as a  function of  $(K-K_{\rm
  RC})$ and fit a linear relation, deriving a representative slope by
  making the average of the single measurements. The left panel  
  of Fig.~\ref{fig:6_7_beta_clusters} shows the
straight  lines defined  by different  clusters, each  in a  different
color.   They    are   consistent    with   a   constant    slope   of
$\beta_{K_{s}}=0.384\pm0.019$ \AA~  mag$^{-1}$, in  excellent agreement
with  those  derived  by  \cite{mauro+2014} using  the  clusters  from
\cite{saviane+2012} combined  with VVV photometry.  This  slope allows
us to define the {\it reduced} EW parameter
\[
W^{\prime}=\Sigma \mathrm{EW} + \beta_{K_{s}}(K-K_{\rm RC})
\]
which is dependent only upon the metallicity of the star.

In order to determine the corresponding  $(K-K_{\rm RC})$ for our bulge stars we 
measure the mean RC magnitude from the complete VVV de-reddened $K_{s}$ 
photometric catalogue corresponding to the observed field, by using the 
reddening map presented in \cite{gonzalez+2011redd}. The 
right panel of Fig.~\ref{fig:6_7_beta_clusters} shows the location
of  our  RC and  RGB  targets,  all in  Baade's  Window,  in the  same
plane. The  color coding traces  metallicity (as measured from  the HR
spectra), and it  is the same used for the  star clusters. Contrary to
what happens in star clusters, bulge  stars in each metallicity bin do
not seem to  define parallel straight lines. We interpret  this as the
result  of the probable disk contamination, the errorbar on the  
metallicity of  each  star, and  the
intrinsic width of the metallicity  bin. Also, while for star clusters
the  x-axis is  a  clean  proxy for  luminosity,  bulge  stars have  a
significant distance  spread, and  the red  clump only  reflects their
mean distance.

\subsection{Reference metallicities: Galactic bulge and globular clusters}

In order to derive an equation converting between $W^{\prime}$ and the
[Fe/H] abundances, a set of independent determinations of [Fe/H] from
high resolution spectra are needed. In what follows we describe how
those were obtained, both for bulge field giants and for a subset of
the globular clusters presented in \cite{warren+2009}. The latter have 
been added to our sample, after showing that they were fully compatible, 
in order to extend the metallicity range on the metal poor side.

\subsubsection{Homogeneous high resolution metallicities determination 
for Galactic bulge stars}

[Fe/H] abundances  for both RC and  RGB stars in Baade's  Window, were
discussed  in  \cite{zoccali+2008}  and \cite{hill+2011}.   They  were
based on GIRAFFE spectra, through  setups HR13 and HR14, at resolution
R$\sim$22,500 and $\sim$28,800,  respectively. After the pre-reduction
and coaddition of the spectra  explained in Section 2, iron abundances
were    measured   by    means   of    individual,   unblended    iron
lines. Conceptually, we used the same method explained in our previous
works,  cited above.  Namely,  first guess  effective temperature  and
gravity  were  derived  from  photometry, and  used  to  construct  an
approximate model  atmosphere. The  latter was  used to  derive [Fe/H]
abundances,  independently,  from  several  FeI and  FeII  lines.  The
stellar   surface  parameters   (T$_{\rm  eff}$,   log  $g$   and  the
microturbulent  velocity   $\chi$)  were  then  refined   by  imposing
consistency  among the  abundances  derived from  different lines.  As
discussed  in  \cite{hill+2011},  the actual  implementation  of  this
method may  yield slightly different final  abundances -especially for
metal rich, cold giants- depending  on which of the stellar parameters
is  fixed first.  In other  words, there  can be  degeneracies in  the
parameter space, due to correlations among the stellar parameters.

In order to  minimize this effect, for the present  work we applied an
automated     procedure,    based     on    the     use    of     GALA
\citep{mucciarelli+2013}.   This  code  is   designed  to  impose  the
excitation  and  ionization equilibria  on  FeI  lines, thus  deriving
T$_{\rm eff}$  and log $g$, respectively,  and to converge on  a value
for the microtubulent  velocity that yields the  same [Fe/H] abundance
for  lines of  different EWs.  The ATLAS9  \citep{castelli+2004} model
atmosphere were used instead of the MARCS ones \citep{gustafsson+2008}
because  GALA is  optimized for  the  former.  The  results are  fully
compatible     with      those     of      \cite{zoccali+2008}     and
\cite{gonzalez+2011redd}.  A  small offset  -within the  errorbars- is
present  between the  resulting metallicities  and those  presented by
\cite{hill+2011}.  The use of GALA,  however, does reduce at least the
statistical errors, as demonstrated  by the exceptionally small spread
of  abundances  in  the  [Mg/Fe]  versus  [Fe/H]  plot,  presented  in
\cite{zoccali+2015} and Gonzalez et al. (2015, {\it in preparation}).

\subsubsection{Globular clusters}

In order to  constrain the CaT versus [Fe/H] calibration  in the metal
poor regime, where bulge stars are sparse, we considered adding to our
sample some  globular clusters. A natural  choice was a subset  of the
\cite{warren+2009} cluster sample. This is  for two main reasons: {\it
  i)} we  had access  to their  spectra, and indeed  we have  shown in
Fig.~\ref{fig:5_comp_ew}  that the  EWs  measured with  our method  is
fully  consistent  with theirs;  and  {\it  ii)} $K_s$  photometry  is
available for all  those stars. In their  work, \cite{warren+2009}
derived a calibration equation between the CaT reduced EWs, as derived
using  the  three   lines,  and  [Fe/H]  measured  in   the  scale  of
\cite{carretta+1997}. The  latter scale, however, has  been revised in
\cite{carretta+2009} (hereafter  C09), to  what has  now become  a new
standard \citep[e.g.][2010 edition]{harris96}.  We therefore updated
the  [Fe/H] abundances  to  the  C09 scale  using  the compilation  in
\cite{carrera+2013}.   For  globular   clusters  without   new  [Fe/H]
measurements,  we  used  instead  the relation  provided  in  C09,  to
transform the metallicity from the old to the new scale. It is worth
emphasizing here that both globular clusters and bulge stars have the
same calcium over iron ratio ([Ca/Fe]$\sim$+0.4) at [Fe/H]$\sim -1$
where they overlap \citep{gonzalez+2011_alphas, carrera+2013}.

The calibration by \cite{warren+2009}  includes several open clusters,
constraining  the  metal  rich  regime. We  excluded  those  from  our
analysis, because we sample the metal rich side with a large number of
bulge  stars,  and  our  goal  is  to  optimize  the  calibration  for
bulge-like giants. In section 4 we will compare our calibration with
open and globular clustr data.

\subsection{The metallicity calibration}
    
\begin{figure*}[ht!]
\centering
{\includegraphics[angle=-90,width=2\columnwidth]{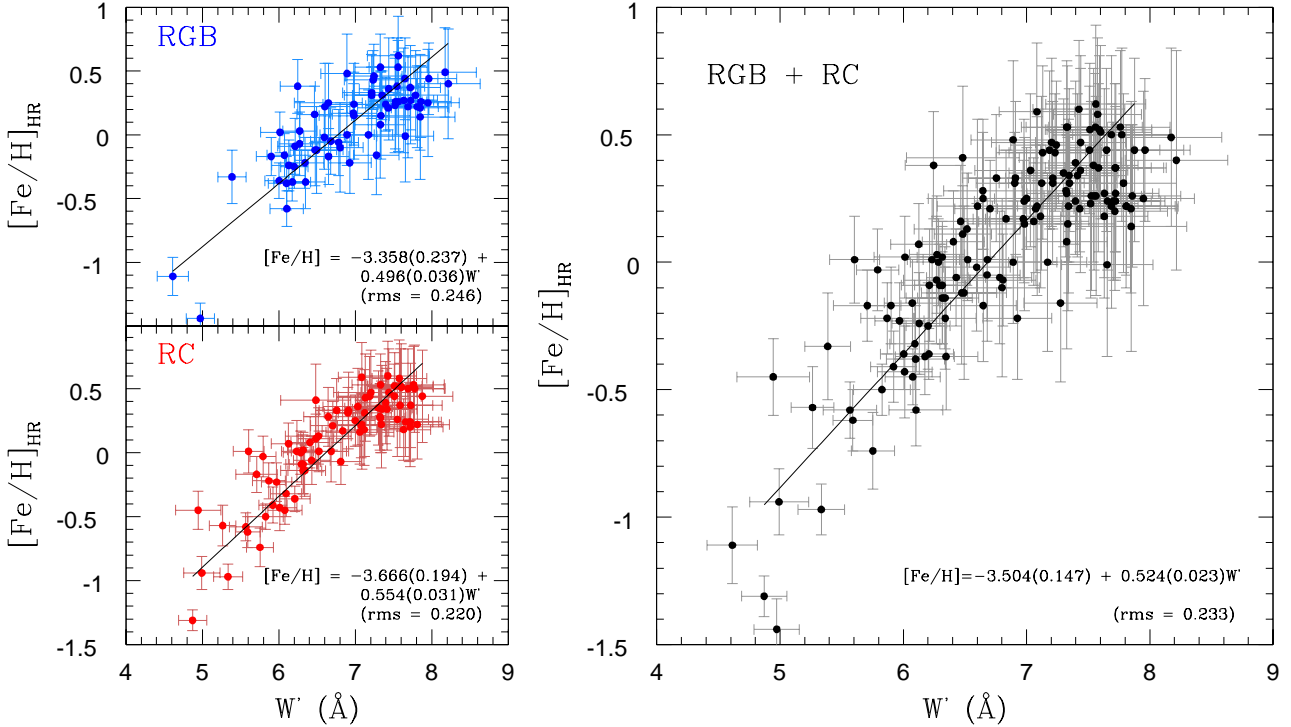}}         
\caption{High resolution metallicities are compared with reduced equivalent 
widths of RGB (blue), RC (red) and the combination of the two (black) bulge 
star samples. The best fitting CaT calibration for each sample is also shown
in each panel.}
\label{fig:7_calib_bulge}
\end{figure*}
    
The calibration  equation between  $W^{\prime}$ and  [Fe/H]$_{\rm HR}$
was derived by  means of an error-weighed least square  fitting of the
data points  relative to individual  stars, both  in the bulge  and in
globular clusters.   This is different  from what is usually  done for
star  clusters,  where the  mean  values  of $\left\langle  W^{\prime}
\right\rangle$ and  $\left\langle \rm  [Fe/H] \right\rangle$  for each
cluster  are  used.   Several  fitting functions  were  tested  before
concluding  that a  linear relation  was the  best choice,  because it
minimizes the artificial {\it  saturation} of the calibration equation
at  high  metallicity   (i.  e.  a  constant   metallicity  at  higher
$W^{\prime}$).  The  latter, indeed, may create  an unphysically sharp
cut-off in the iron distribution function.

As a first step, a linear fitting was performed between $W^{\prime}$ and  
[Fe/H]$_{\rm HR}$ for RGB and RC stars, independently. The results are
shown in the left panels of Fig.~\ref{fig:7_calib_bulge}, together with
the rms ($\sim$0.23) which is comparable to the uncertainties of the
[Fe/H]$_{\rm HR}$ measurements. The fits are consistent with each other,
and can therefore be combined in the single linear relation shown in
the right panel of the same figure, namely:
\begin{equation} \label{Fe_bulge}
\mathrm{[Fe/H]_{RGB+RC}}=(-3.504 \pm 0.147) + (0.524 \pm 0.023)W^{\prime} 
\end{equation}
whith a rms of $0.233$ dex.

As a second step, 7 globular clusters were added to the plot, covering
the metallicity range $-2.3<$[Fe/H]$<-0.7$. The fit of a linear
relation between the mean $W^{\prime}$ and mean [Fe/H], for the
clusters only, yields:
\begin{equation} \label{Fe_gc}
\mathrm{[Fe/H]_{GC}}=(-3.175 \pm 0.094) + (0.462 \pm 0.025)W^{\prime} 
\end{equation}
which, though slightly different from the equation found for the bulge
(Eq.~\ref{Fe_bulge}), shows  a smooth transition between  the two when
they overlap.  This suggests  that a single  calibration can  be used,
valid across  the whole  range $-2.3<$[Fe/H]$<+0.7$.  Nevertheless, as
globular cluster  and Galactic bulge stars  do not have the  same data
sampling  (i.e.    for  globular  clusters   $\left\langle  W^{\prime}
\right\rangle$  and  $<$[Fe/H]$_{\rm HR}>$  are  used  instead of  the
measurements for individual stars) we cannot directly combine the data
sets in  order to  fit a  single function,  because the  resulting fit
would be  dominated by the metal  poor data, having smaller  errors in
[Fe/H].   Because of  this, we  decided to  create two  evenly sampled
fiducial  data  sets taken  from  the  equations (\ref{Fe_bulge})  and
(\ref{Fe_gc}), in  the $W^{\prime}$  range between  $0 <  W^{\prime} <
5.4$ and  $5.2 < W^{\prime} <  10$ respectively. These data  sets were
then combined into a single one  and fitted by a second order function
in order  to take into account  the small change of  slope between the
two linear equations (Eq.~\ref{Fe_bulge} and \ref{Fe_gc}).

The resulting fit gives:
\begin{equation} \label{CaT_cal}
\mathrm{[Fe/H]_{CaT}} = -3.150 + 0.432W^{\prime} + 0.006W^{\prime 2}
\end{equation}
with a rms dispersion around the fit of $0.197$ dex. 
No errors are quoted for the coefficients here, because, due to the procedure
adopted to combine globular clusters and bulge stars, this is more a
{\it fiducial} relation, characterizing the locus of the data in the
$W^{\prime}$ versus [Fe/H] plane, rather than a formal fit.

\begin{figure}
\centering
{\includegraphics[angle=0,width=1.0\columnwidth]{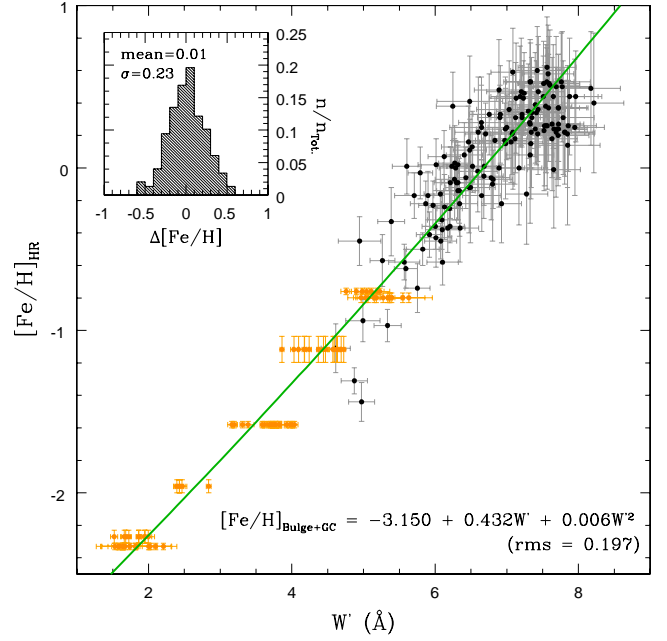}}         
\caption{The CaT calibration for the combination between bulge stars (black) with 
globular clusters from \cite{warren+2009} (orange). The upper left panel 
shows the corresponding calibration metallicity residuals for bulge stars.}
\label{fig:9_calib_bulge_gc}
\end{figure}

When applying  the calibration,  errors on  the derived  [Fe/H] values
will  be calculated  as the  quadratic sum  of the  errors on  the EW,
propagated  through equation~\ref{CaT_cal},  and the  rms of  the same
equation.   Figure~\ref{fig:9_calib_bulge_gc} summarizes  our results,
showing the final CaT vs [Fe/H] calibration, as a green line, together
with the  bulge stars (black  dots) and the globular  clusters (orange
dots).   For   the  latter,   individual  $W^{\prime}$   -measured  by
\cite{warren+2009}- are plotted for individual  stars, all at the same
cluster  $<$[Fe/H]$_{\rm   HR}$.  The  upper-left  corner   shows  the
histogram  of the  residuals for  bulge stars,  i.e., the  differences
between [Fe/H]$_{\rm HR}$ and [Fe/H]$_{\rm CaT}$.  The distribution is
symmetric around  zero and the  dispersion is consistent with  the rms
quoted above.

\section{Comparison with previous calibrations}

\begin{figure}
\centering
{\includegraphics[angle=0,width=1.0\columnwidth]{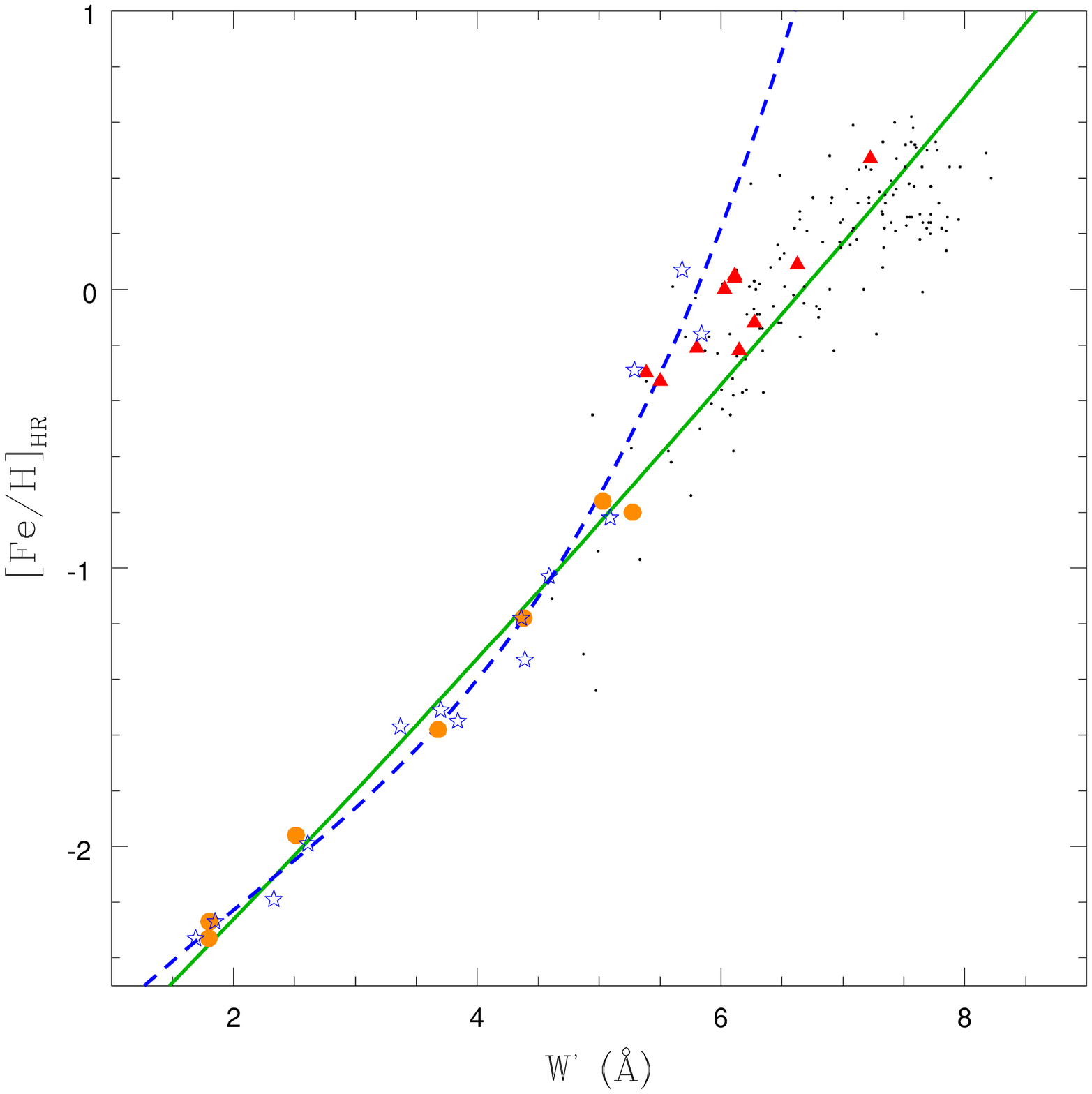}}         
\caption{High resolution metallicity versus the reduced equivalent width 
for different star samples: bulge stars (black dots), globular (orange circles) 
and open (red triangles) clusters from \cite{warren+2009}, and globular 
clusters (blue stars) from Saviane et al. \citep{saviane+2012}. Our 
CaT calibration is plotted as a green 
line, while the cubic calibration from \cite{saviane+2012} is the dashed
blue line.}
\label{fig:11_calib_comp}
\end{figure}

Different CaT  vs [Fe/H] calibration  have been proposed in  the past,
differing among each others mainly because of the different methods to
measure EWs, and the different  [Fe/H] scales. A recent calibration is
presented  in \cite{saviane+2012}  who  used 14  globular clusters  to
define a  cubic equation covering the  range $-2.33<$[Fe/H]$<0.07$, in
the C09 scale. Since the EWs  measured by us are fully compatible with
the  \cite{saviane+2012}  (c.f.,  Fig.~\ref{fig:5_comp_ew}),  the  two
calibrations can also be directly compared.

Figure~\ref{fig:11_calib_comp}       is      color       coded      as
Fig.~\ref{fig:9_calib_bulge_gc}, with bulge giants as small black dots
and globular  and open  cluster mean  values as  orange and  red dots,
respectively \citep{warren+2009}. Our calibration  is shown as a green
line.   Also   shown  are   the  globular  clusters   in  \cite[][blue
  stars]{saviane+2012}  and  their  calibration equation,  as  a  blue
dashed line.  The two calibrations, as  well as the original  one from
\cite{warren+2009}, are  consistent with each other  for [Fe/H]$<-0.8$
dex.  The higher order fit  in \cite{saviane+2012} is required only by
the three  most metal rich  globular clusters in their  study, namely:
NGC~5927  ([Fe/H]=$-0.29$),  NGC~6553  ([Fe/H]=$-0.16$)  and  NGC~6528
([Fe/H]=$+0.07$).  Interestingly,  the same behaviour is  seen in C09,
where  the updated  \cite{rutledge+1997}  calibration  changes from  a
linear to  a cubic fit when  NGC~6553 is added to  the original sample
(see  Fig.  A.2  from  \cite{carretta+2009}).  Moreover,  as noted  by
Saviane et al., NGC~6528 has  $W^{\prime}$ lower than NGC~6553, though
being more metal rich.

Although we are  not able to explain the difference  between the three
most metal  rich globular clusters and  bulge stars, we note  that the
cubic relation  of \cite{saviane+2012} is clearly  not appropriate for
bulge stars, which are the main  focus of our interest.  Also shown in
Fig.~\ref{fig:11_calib_comp}    are    the    open    clusters    from
\cite{warren+2009}. Although  the authors originally derived  a linear
calibration, the  use of  the updated [Fe/H]  from \cite{carrera+2013}
for the open clusters is not compatible anymore with a single straight
line.   The  difference between  orange  dots  and red  triangles,  in
Fig.~\ref{fig:11_calib_comp},  looks more  like  a systematic  shift,
rather than a change of slope.  A similar shift can be observed in the
[Ca/H] versus [Fe/H] plot shown in Fig.~11 of \cite{carrera+2013}, and
it might be ascribed to inhomogeneities in the [Fe/H] sources for open
clusters, or to  a systematic effect due to age,  compared to globular
clusters.


\section{Conclusions}

Using RGB and  RC stars from Baade's window we  have constructed a CaT
versus  [Fe/H] calibration,  valid  in  the metal  rich  regime up  to
[Fe/H]=+0.7. Since  bulge stars are rather  sparse for [Fe/H]$<-0.5$,
we  used  globular  clusters  to  constrain  the  behaviour  of  the
  calibration  relation in  the metal  poor  regime. We  found that  a
  single  linear  relation  can  be  fitted  across  the  whole  range
  $-2.3<$[Fe/H]$<+0.7$. A  minor second order deviation  from a single
  linear relation,  however, gives a  better fit, properly  taking into
  account  the  small  difference  between the  linear  best  fit  for
  globular  clusters (equation  \ref{Fe_bulge}) and  for bulge  giants
  (equation \ref{Fe_gc}). It is worth emphasizing that both 
  globular clusters and bulge stars have 
[Ca/Fe]$\sim+0.3$ in the common metallicity range, therefore a possible
dependence of the calibration upon the [Ca/Fe] ratio would not affect the
combination of the two datasets. At higher metallicities the bulge [Ca/Fe]
drops to zero \citep{gonzalez+2011,gonzalez+2015} but the CaT-vs-Fe
calibration found here is still almost linear, possibly due to the combined
effect of the strong saturation of CaT lines. The 
new calibration, represented by equation
  \ref{CaT_cal}, allows one to derive the iron content of metal rich stars by
  means of low-resolution spectra.  The latter are the only viable way
  to sample a  large number of stars in high  extinction regions, such
  as the  Galactic bulge, or,  in the  future, in metal  rich external
  spheroid such as the inner halo of M31.


\begin{acknowledgements}
We  gratefully  acknowledge  support   by  the  Ministry  of  Economy,
Development, and Tourism's Millennium Science Initiative through grant
IC120009, awarded  to The Millennium Institute  of Astrophysics (MAS),
by  Fondecyt  Regular  1110393 and 1150345  and   by  the  BASAL-CATA  Center  for
Astrophysics and Associated Technologies PFB-06. SV acknowledges to the 
Henri Poincare Junior Fellowship and the Chilean French embassy doctoral 
internship program to support research travels to the Nice Observatory 
during 2011 and 2013.
\end{acknowledgements}

\bibliographystyle{aa}
\bibliography{mybiblio}


\end{document}